\documentclass[prx,aps,twocolumn,amsmath,amssymb]{revtex4-1}
\pdfoutput=1
\usepackage{hyperref}
\usepackage{graphicx}

\def\eqq#1{Eq.~(\ref{#1})}
\def\eq#1{(\ref{#1})}
\def\av#1{\langle #1 \rangle}
\def\f#1{Fig.~\ref{#1}}

\def\c#1{~\cite{#1}}

\def\ed{\epsilon_{\rm d}}
\def\es{\epsilon_{\rm s}}

\def\gb{\gamma_{\rm b}}
\def\gr{\gamma_{\rm r}}

\def\Gb{\Gamma_{\rm b}}
\def\Gr{\Gamma_{\rm r}}

\def\lb{\lambda_{\rm b}}
\def\lr{\lambda_{\rm r}}

\def\kt{k_{\rm B}T}
\def\e{{\rm e}}

\def\beq{\begin{equation}}
\def\eeq{\end{equation}}
\def\bea{\begin{eqnarray}}
\def\eea{\end{eqnarray}}

\begin{document}

\title{Similarity of ensembles of trajectories of reversible and irreversible growth processes}
\author{Katherine Klymko$^1$}
\author{Juan P. Garrahan$^2$}
\author{Stephen Whitelam$^3$}
\email[]{swhitelam@lbl.gov}

\affiliation{$^1$Department of Chemistry, University of California at Berkeley, Berkeley, CA 94720, USA \\
$^2$School of Physics and Astronomy, University of Nottingham, Nottingham NG7 2RD, UK\\
$^3$Molecular Foundry, Lawrence Berkeley National Laboratory, 1 Cyclotron Road, Berkeley, CA 94720, USA}

\begin{abstract}
Models of bacterial growth tend to be `irreversible', allowing for the number of bacteria in a colony to increase but not to decrease. By contrast, models of molecular self-assembly are usually `reversible', allowing for addition and removal of particles to a structure. Such processes differ in a fundamental way because only reversible processes possess an equilibrium. Here we show at mean-field level that dynamic trajectories of reversible and irreversible growth processes are similar in that both feel the influence of attractors, at which growth proceeds without limit but the intensive properties of the system are invariant. Attractors of both processes undergo nonequilibrium phase transitions as model parameters are varied, suggesting a unified way of describing reversible and irreversible growth. We also establish a connection at mean-field level between an irreversible model of growth (the magnetic Eden model) and the equilibrium Ising model, supporting the findings made by other authors using numerical simulations.
\end{abstract}

\maketitle

{\em Introduction.} Physical growth processes can be reversible, allowing for the number of particles present in a system to increase and decrease, or irreversible, allowing only for an increase of particle number. For example, bacterial colony growth is usually considered to be irreversible, because bacteria multiply but do not disappear\c{eden1961two,mazzitello2015far}. By contrast, models of molecular self-assembly are usually reversible, allowing for particle attachment and detachment\c{hagan2006dynamic,wilber2007reversible,rapaport2008role,drossel1997model}. The two types of process are fundamentally different in that reversible processes possess an equilibrium at which growth ceases, while irreversible processes do not. Usually one of these processes is chosen to model a particular physical system, and so comparison between the two is rarely made. Here we compare reversible and irreversible stochastic growth processes in a mean-field (space independent) setting. We show that despite their differences in respect of equilibrium, the two types of process can display similar behavior when growth is allowed to proceed without limit. Specifically, ensembles of dynamic trajectories are governed by attractors in phase space at which the averaged properties of the system, scaled by system size, are invariant. These attractors undergo nonequilibrium phase transitions as model parameters are varied, suggesting a unified way of describing reversible and irreversible processes. For one particular irreversible process, a mean-field version of the magnetic Eden model (MEM)\c{eden1961two,candia2008magnetic,candia2001comparative}, we show that its nonequilibrium phase behavior is that of the mean-field {\em equilibrium} Ising model. This finding provides additional evidence for a ``nontrivial correspondence between the MEM for the irreversible growth of spins and the equilibrium Ising model'' (in distinct spatial dimensions) conjectured by other authors on the basis of numerical simulations\c{candia2008magnetic,candia2001comparative}.

{\em Modeling reversible and irreversible growth.} We consider reversible and irreversible stochastic growth processes in the simplest limit, ignoring spatial degrees of freedom and resolving only the numbers of particles in the system. By `reversible' we mean simply that particles may enter {\em and} leave the system, and we intentionally do not require that rates are derived from the principle of detailed balance. We consider growth of a system composed of two types of particle, labeled `red' and `blue'. The state of the system is defined at any instant by the number of red particles $r$ and blue particles $b$ it contains, or equivalently by the system's `size' $N\equiv b+r$ and `magnetization' $m \equiv (b-r)/(b+r)$. We add blue particles to the system with rate $\lb$, and red particles with rate $\lr$. We remove blue and red particles from the system with respective rates $\gb$ and $\gr$. For an irreversible process these latter two rates are zero. We allow rates to depend on the instantaneous magnetization of the system but not (directly) its size. We impose this requirement in order to model a notional growth process in which rates of particle addition and removal to a structure scale with the size of the interface between the structure and its environment. We then assume the limit of a large structure whose interfacial area does not change appreciably during the growth process, and we divide addition and removal rates by the (constant) surface area in order to obtain the rates stated above. 

We studied this class of growth processes using a continuous-time Monte Carlo protocol\c{gillespie2005general}. To interpret these simulations we derived a set of analytic expressions for averages over dynamic trajectories, in the limit of vanishing particle-number fluctuations (see Sections~\ref{supp1}--\ref{supp3}). Under these conditions the net rates of addition of blue and red particles are $\Gb(m) = \lb-\gb$ and $\Gr(m) = \lr-\gr$. The time evolution of system size is $\dot{N} = \Gb+\Gr$. The requirement for equilibrium, by which we mean a zero-growth-rate stationary solution, is $\Gb(m_0) = 0 = \Gr(m_0)$, where $m_0$ is the magnetization of the system in equilibrium. These relations can be satisfied by a reversible process but not an irreversible one, except in the trivial limit of zero addition rate. Thus only a reversible process has an equilibrium for which $\dot{N}= 0 = \dot{m}$. However, both types of process admit a steady-state growth regime for which $\dot{N}>0$ and $\dot{m}=0$. The time evolution of magnetization is $\dot{m} = N^{-1} \left[ \Gb-\Gr-m(\Gb+\Gr) \right]$, which vanishes for $m=m_\star$ such that
\beq
\label{ss}
m_\star = \frac{\Gb(m_\star) - \Gr(m_\star)}{\Gb(m_\star)+\Gr(m_\star)}.
\eeq
Thus there exist nullclines, at which $\dot{m}=0$, in the space of dynamic growth trajectories. The existence of such nullclines requires only that {\em net} rates of particle addition are positive, whether or not removal rates vanish, and so can be displayed by reversible {\em and} irreversible processes. We shall show that these nullclines can be attractors, stable with respect to perturbations, and so constitute fixed lines to which dynamic trajectories flow. Furthermore, these attractors undergo nonequilibrium phase transitions as model parameters are varied.

We now specialize the discussion to two models that might be regarded as reversibly- and irreversibly growing versions of the mean-field Ising model. The irreversible stochastic process we consider is a space-independent version of the magnetic Eden model\c{eden1961two,candia2008magnetic,candia2001comparative}, closely related to a model studied in Ref.\c{morris2014growth}. Addition of red and blue particles occurs with rates that are Arrhenius-like in the energy of interaction between a single particle and the system, $\lr =\frac{1}{2} \e^{-m J-h}$ and $\lb = \frac{1}{2}\e^{m J+h}$. Here $J$ is the interparticle coupling and $h$ a magnetic field (we set $m=0$ when $N=0$). We allow no particle removals, setting $\gb=\gr=0$. There is therefore no equilibrium. The analytic evolution equations, averaged over trajectories, read $\dot{N} = \cosh(m J +  h)$ and
\beq
\label{mdot_eden}
\dot{m} = N^{-1} \left[ \sinh(m J + h) -m \cosh(m J +  h)\right],
\eeq
and admit the nullcline
\beq
\label{eos_eden}
m_\star = \tanh(m_\star J +  h).
\eeq
This equation is equivalent to the well-known mean-field expression for Ising model thermodynamics\c{chandler, binney1992theory}. For $h=0$, Equations~\eq{mdot_eden} and \eq{eos_eden} indicate that the nullcline $m_\star=0$ is an attractor for $J \leq J_{\rm c} = 1$ and a repeller for $J>1$. For $J>1$ two symmetric attractors emerge as ${m_\star}_\pm \sim \pm (J-1)^{1/2}$. In other words, these equations describe a continuous phase transition of dynamic trajectories that are `unmagnetized' for  $J<1$ and `magnetized' for $J>1$, via a critical point at $J=1$. Thus, at mean-field level, nonequilibrium trajectories of the magnetic Eden model possess phase behavior identical to that of the equilibrium Ising model~\footnote{The same conclusion follows from the variant of Eq. (14) of Ref.\c{morris2014growth} that would be obtained by setting, in Eq. (5) of that reference, the spin-flip terms $f$ to zero and requiring the vanishing of the remaining term.}. This result provides an analytic connection between models supporting the findings of Refs\c{candia2008magnetic,candia2001comparative}, which demonstrated a numerical equivalence between phase transitions, in distinct spatial dimensions, of Eden and Ising models (see also\c{mazzitello2015far}). This result also appears to be consistent with general arguments suggesting that probabilistic irreversible cellular automata with Ising-like symmetry lie in the universality class of the equilibrium Ising model\c{grinstein1985statistical} (see also\c{bar1990structure,drossel1997model}).
\begin{figure}
   \centering
\includegraphics[width=0.8\linewidth]{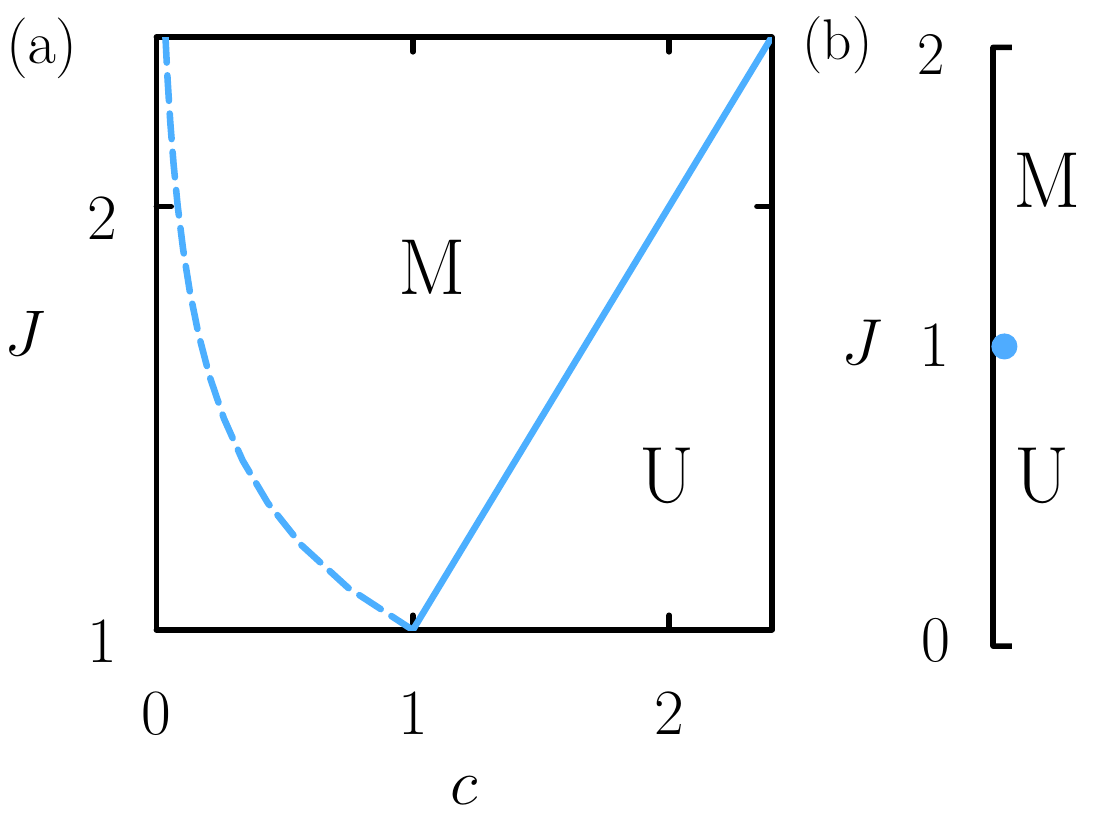} 
   \caption{Phase diagrams of the stable dynamic attractors of the reversible (a) and irreversible (b) models of growth, for $h=0$. U and M indicate unmagnetized and magnetized regions, respectively, with the latter admitting two stable attractors. The solid line in (a) and dot in (b) denote continuous phase transitions. To the left of the dotted line in (a) we have no growth.}
   \label{fig1}
\end{figure}
\begin{figure}
   \centering
   \includegraphics[width=\linewidth]{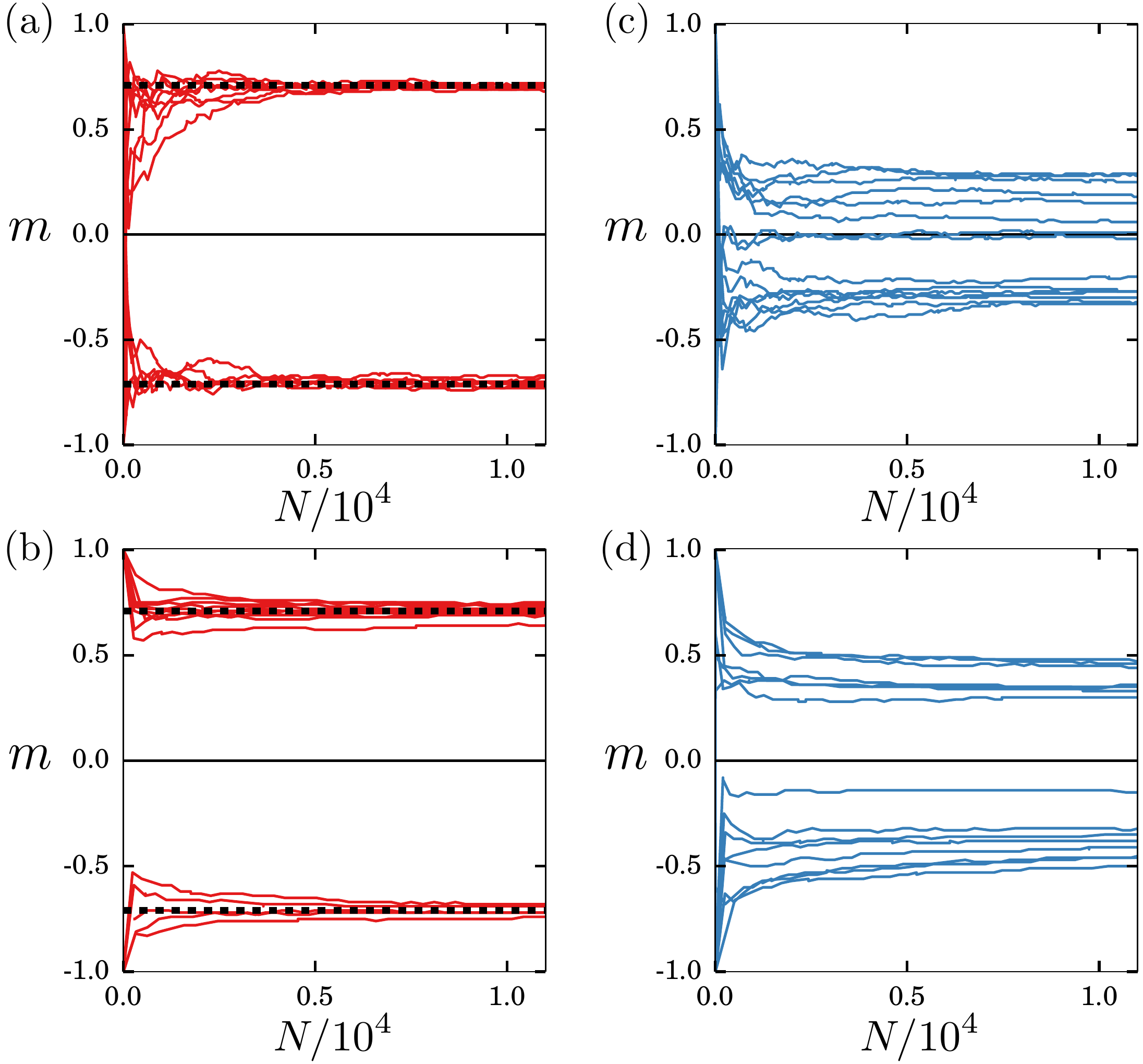} 
   \caption{Dynamic growth trajectories in the space of magnetization $m$ and system size $N$ for reversible (top) and irreversible (bottom) models. We show trajectories in the magnetized regime (a,b) with attractors marked as dotted lines, and at a dynamic critical point (c,d) where the attractor lies at zero magnetization. Parameters: (a) $J=2.5$ and $c=2$, (b) $J=1.25$, (c) $J=2$ and $c=2$, (d) $J=1$.}
   \label{fig2}
\end{figure}

\begin{figure*}
   \includegraphics[width=0.75\linewidth]{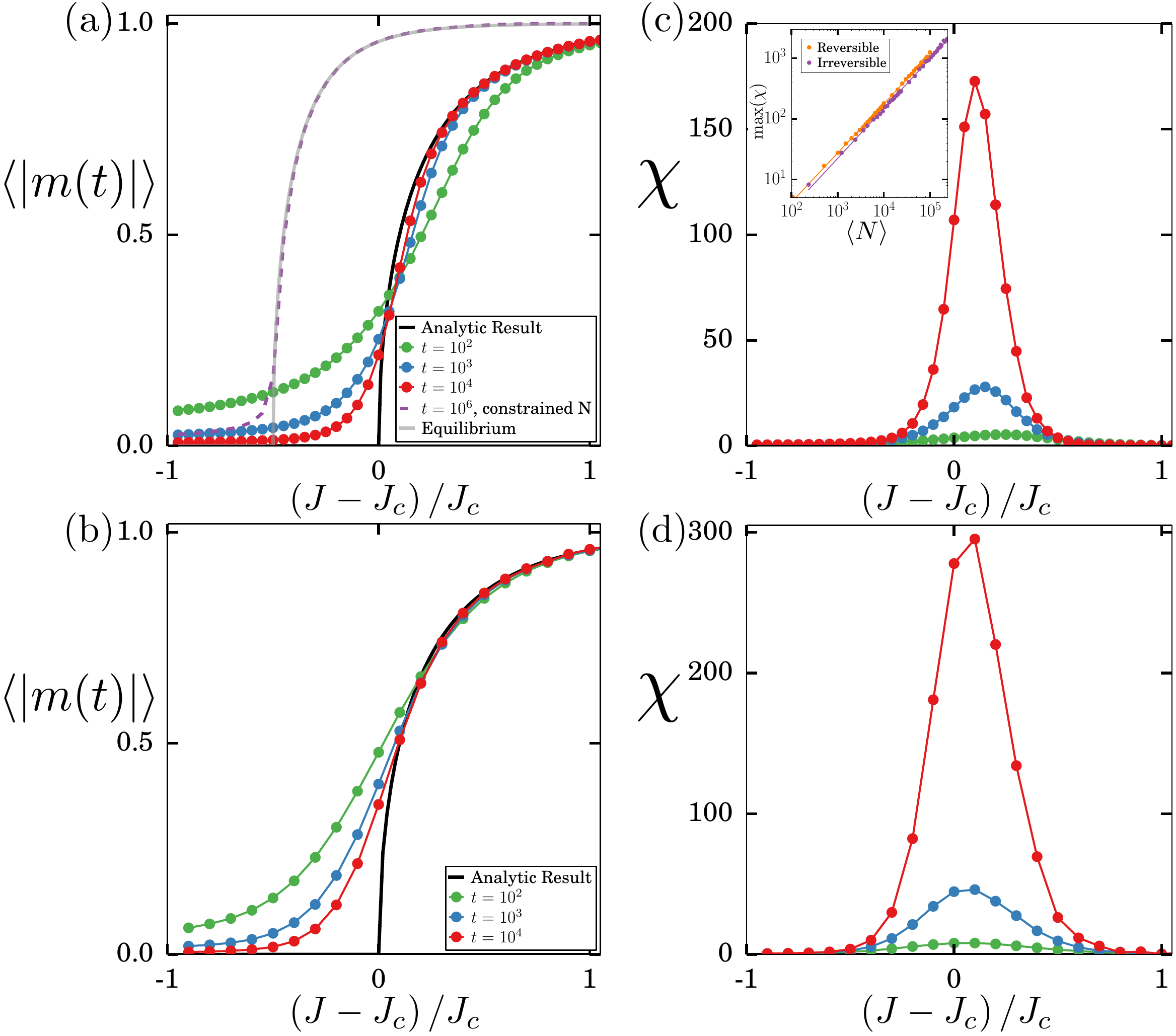} 
   \caption{Trajectory-to-trajectory averages (a,b) and variance (c,d) of magnetization taken over ensembles of $10^5$ trajectories, at various values of $J$, for (a,c) reversible ($c=2$) and (b,d) irreversible growth. The two types of process display similar phase transitions. Numerical simulations are overlaid on the analytic results \eq{eos_rev} (a) and \eq{eos_eden} (b). In panel (a) we also show the results of simulations done in the presence of a system size constraint, overlaid on \eq{rev_3} (see \f{fig4}).}
   \label{fig3}
\end{figure*}

The reversible model we consider is the stochastic process whose fluctuation-free limit is described in Refs.\c{whitelam2014self,whitelam2015minimal}. We assume constant rates of particle addition, $\lb = p c $ and $\lr = (1-p) c$, where $c$ is a notional `solution' concentration and $p$ is the fraction of particles in solution that are blue. To make contact with Ising model nomenclature we introduce the magnetic field $h$ via $p \equiv \e^h/(2 \cosh h)$. Unbinding rates are Arrhenius-like, appropriate for particles escaping from a structure via thermal fluctuations, and are $\gb=\frac{1}{2} \e^{-m J} (1+m)$ and $\gr=\frac{1}{2} \e^{m J} (1-m)$ (supplemented by the restriction that particle numbers cannot be negative). Note that these rates are intentionally not designed to satisfy detailed balance with respect to a particular energy function; however, the process still possesses an equilibrium. The fluctuation-free evolution equations are $\dot{N} = c-  \cosh(m J) +m  \sinh (m J)$ and 
\beq
\label{rev_2}
\dot{m} = N^{-1} \left[ \left(1-m^2\right) \sinh(m J) - c \left(m-\tanh h\right)\right].
\eeq
Equilibrium is achieved when $c_0^2 = \left( 1-m_0^2\right) \cosh^2 h$, with 
\beq
\label{rev_3}
m_0 = \tanh(m_0 J+h).
\eeq
Thus the {\em equilibrium} phase behavior of this model is identical to the {\em nonequilibrium} phase behavior \eq{eos_eden} of the irreversible model, and to the equilibrium phase behavior of the mean-field Ising model.

Persistently-growing trajectories admit the nullcline
\beq
\label{eos_rev}
c\left(m_\star -\tanh h\right)=\left( 1-(m_\star)^2\right) \sinh (m_\star J).
\eeq
This nullcline is different in detail to that of the irreversible model, \eqq{eos_eden}. However, \eq{eos_rev} and \eq{rev_2} describe, for $J < \sqrt{6}$, a similar nonequilibrium continuous phase transition between unmagnetized and magnetized trajectories. The `magnetic' critical exponent is $1/2$, as for the irreversible case, i.e. magnetization emerges for $J>J_{\rm c}$ as ${m_\star}_\pm \sim \pm (J-J_{\rm c})^{1/2}$, where $J_{\rm c}=c$ is the critical point of the reversible process.

In \f{fig1} we show the nonequilibrium phase diagrams derived from \eq{mdot_eden}, \eq{eos_eden}, \eq{rev_2} and \eq{eos_rev}. These diagrams indicate the nature of the dynamic attractors $m_\star$ in a space of model parameters: in regions U and M the stable attractors possess zero and nonzero magnetization, respectively. Both models exhibit phase transitions at which the nature of the attractors changes qualitatively.

Numerical simulations accommodating particle-number fluctuations confirm these analytic expectations, and provide additional insight into the nature of phase transitions of ensembles of trajectories. We began simulations (in general) with no particles present, and advanced time by an amount $1/(\gr+\gb+\lr+\lb)$ following every Monte Carlo move. In \f{fig2} we show that dynamic trajectories feel the influence of the dynamic attractors predicted analytically. In the magnetized region trajectories `flow' to one of the two stable magnetized attractors, while at the critical point the stable attractor is unmagnetized. {\em Individual} trajectories fluctuate increasingly slowly in $m$-space as $N$ increases (because, for large $N$, fluctuations of particle number change magnetization by an amount $\propto N^{-1}$), and so e.g. the likelihood of a magnetized trajectory spontaneously reversing its magnetization becomes vanishingly small (see Ref.\c{morris2014growth}). However, {\em ensembles} of trajectories show behavior characteristic of a phase transition. In \f{fig3}(a,b) we show averages of $|m(t)|$ over $10^5$ dynamic trajectories generated at several different values of $J$. We define averages of an observable $A(t)$ as $\av{A(t)} \equiv K^{-1} \sum_{i=1}^K A_i(t)$, where $A_i(t)$ is the value of $A(t)$ for the $i$th of $K$ trajectories. Trajectory averages evolve as $t$ increases toward the attractor. This evolution is in general slow, because the mobility of individual trajectories is low: ignoring fluctuations we expect small departures $\av{\delta m(t)}$ from the attractor to decay -- above, at, and just below the critical point --  as $t^{-q}$, $(\ln t)^{-1/2}$, and $t^{2q}$, respectively, where $q=1-J$ for the irreversible process and $q=(c-J)/(c-1)$ for the reversible one. 
\begin{figure}
   \centering
 \includegraphics[width=0.9\linewidth]{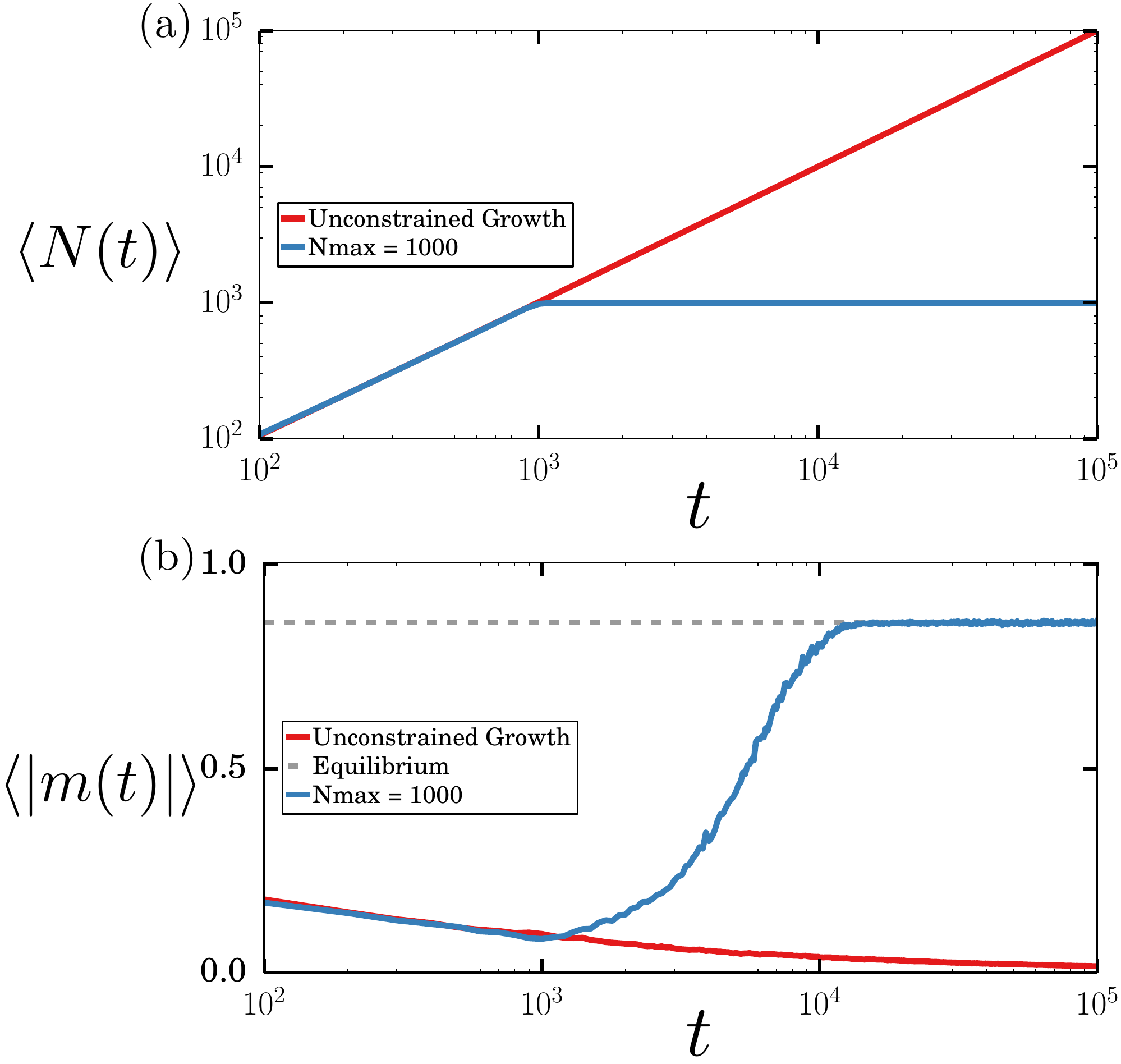}
   \caption{Evolution of $\av{N}$ (a) and $\av{|m|}$ (b) for size-limited reversible growth shows that trajectories fall under the influence of the dynamic attractor while growth persists, and evolve to the equilibrium attractor when the size constraint is reached. Lines denote averages over 500 trajectories. Parameters: $J=1.5,c=2$. The line labeled `constrained' in \f{fig3}(a) shows the results of similarly constrained simulations for several values of $J$.}
   \label{fig4}
\end{figure}

Trajectory-to-trajectory {\em fluctuations}, which are neglected by our analytic expressions, also show behavior characteristic of a phase transition. In \f{fig3}(c,d) we show the trajectory-to-trajectory fluctuations of magnetization, $\chi \equiv \av{N(t)} (\av{m^2(t)} - \av{|m(t)|^2})$ (the quantity ${\rm var}(|M|)/\av{N}$, where $M=mN$, behaves similarly). For both models $\chi$ displays at the critical point a peak that increases in height with average system size as $\av{N(t)}^{0.82(1)}$ over the time interval shown (see inset to \f{fig3}(b)). While individual trajectories flow to stable attractors as time increases, thereby causing ${\rm var} (|m|)$ to decrease with time, the same trajectories also acquire more particles, and the combination $\av{N}\, {\rm var}(|m|)$ {\em increases} with time over the interval simulated. Such `sharpening' of a phase transition with increasing time is reminiscent of behavior seen in glassy models that display phase transitions in space-time\c{hedges2009dynamic}. In the asymptotic limit (when $N\to \infty$ and $m =m_\star$ is constant) we expect the evolution of $M=mN$ to resemble that of a random walker, and so ${\rm var}(|M|) \propto t$. Thus ensembles of trajectories feel the effect of dynamic attractors, but can in certain regimes of phase space take considerable time to reach them.

In some limits the two types of process can be clearly distinguished. All growth processes must eventually end. A bacterial colony will run out of food, and a self-assembled structure will come to equilibrium with its environment. In this limit the difference between reversible and irreversible processes becomes apparent. In \f{fig4} we show dynamic simulations of the reversible model carried out in the presence of an additional rule that forbids any addition that would cause the system to contain more than $10^3$ particles. During the growth phase dynamic trajectories fall under the influence of the dynamic attractor, but when the system size limit is reached trajectories evolve to an attractor similar to that of the equilibrium one; see also the line labeled `constrained' in \f{fig3}(a). Trajectories of the irreversible model, in the presence of a size constraint, simply cease to evolve. The behavior of the reversible model gives insight into the behavior of the lattice models of growth of Refs.\c{whitelam2014self,whitelam2015minimal}. These models obey detailed balance, and so must eventually evolve to equilibrium, but during a period of growth they display nonequilibrium behavior consistent with that of a persistently-growing process. The present results indicate that one can consider dynamic trajectories of a reversible growth process to feel the effect of both nonequilibrium and equilibrium attractors, the relative influence of which varies over the lifetime of the trajectory.

{\em Conclusions.}  We have shown at mean-field level that reversible and irreversible growth processes are similar in that both admit attractors in the space of dynamical trajectories. At these attractors growth proceeds without limit, but averaged intensive properties of the system are time-invariant. Attractors of both types of process can undergo similar nonequilibrium phase transitions. We have also established a connection at mean-field level between an irreversible model of growth (the magnetic Eden model) and the equilibrium Ising model, supporting the findings made by other authors using numerical simulations. There is sustained interest in nucleation and growth pathways of molecular\c{sear2007nucleation}, active\c{redner2016classical} and living\c{eden1961two} matter. Our results indicate that certain qualitative properties of nonequilibrium growth trajectories are insensitive to details of the microscopic transition rates that produce them, suggesting a unified way of describing growth processes of distinct microscopic entities. 

{\em Acknowledgements.} We thank the organizers of the EPSRC workshop `Collective Behaviour in Growing Systems', Bath University, Nov 2014, which motivated this work. KK acknowledges support from an NSF Graduate Research Fellowship. JPG was supported by EPSRC Grant No. EP/K01773X/1. This work was done as part of a User project at the Molecular Foundry at Lawrence Berkeley National Laboratory, supported by the Office of Science, Office of Basic Energy Sciences, of the U.S. Department of Energy under Contract No. DE-AC02--05CH11231.


\begin{thebibliography}{20}%
\makeatletter
\providecommand \@ifxundefined [1]{%
 \@ifx{#1\undefined}
}%
\providecommand \@ifnum [1]{%
 \ifnum #1\expandafter \@firstoftwo
 \else \expandafter \@secondoftwo
 \fi
}%
\providecommand \@ifx [1]{%
 \ifx #1\expandafter \@firstoftwo
 \else \expandafter \@secondoftwo
 \fi
}%
\providecommand \natexlab [1]{#1}%
\providecommand \enquote  [1]{``#1''}%
\providecommand \bibnamefont  [1]{#1}%
\providecommand \bibfnamefont [1]{#1}%
\providecommand \citenamefont [1]{#1}%
\providecommand \href@noop [0]{\@secondoftwo}%
\providecommand \href [0]{\begingroup \@sanitize@url \@href}%
\providecommand \@href[1]{\@@startlink{#1}\@@href}%
\providecommand \@@href[1]{\endgroup#1\@@endlink}%
\providecommand \@sanitize@url [0]{\catcode `\\12\catcode `\$12\catcode
  `\&12\catcode `\#12\catcode `\^12\catcode `\_12\catcode `\%12\relax}%
\providecommand \@@startlink[1]{}%
\providecommand \@@endlink[0]{}%
\providecommand \url  [0]{\begingroup\@sanitize@url \@url }%
\providecommand \@url [1]{\endgroup\@href {#1}{\urlprefix }}%
\providecommand \urlprefix  [0]{URL }%
\providecommand \Eprint [0]{\href }%
\providecommand \doibase [0]{http://dx.doi.org/}%
\providecommand \selectlanguage [0]{\@gobble}%
\providecommand \bibinfo  [0]{\@secondoftwo}%
\providecommand \bibfield  [0]{\@secondoftwo}%
\providecommand \translation [1]{[#1]}%
\providecommand \BibitemOpen [0]{}%
\providecommand \bibitemStop [0]{}%
\providecommand \bibitemNoStop [0]{.\EOS\space}%
\providecommand \EOS [0]{\spacefactor3000\relax}%
\providecommand \BibitemShut  [1]{\csname bibitem#1\endcsname}%
\let\auto@bib@innerbib\@empty
\bibitem [{\citenamefont {Eden}(1961)}]{eden1961two}%
  \BibitemOpen
  \bibfield  {author} {\bibinfo {author} {\bibfnamefont {M.}~\bibnamefont
  {Eden}},\ }\href@noop {} {\bibfield  {journal} {\bibinfo  {journal} {Dynamics
  of fractal surfaces}\ }\textbf {\bibinfo {volume} {4}},\ \bibinfo {pages}
  {223} (\bibinfo {year} {1961})}\BibitemShut {NoStop}%
\bibitem [{\citenamefont {Mazzitello}\ \emph {et~al.}(2015)\citenamefont
  {Mazzitello}, \citenamefont {Candia},\ and\ \citenamefont
  {Albano}}]{mazzitello2015far}%
  \BibitemOpen
  \bibfield  {author} {\bibinfo {author} {\bibfnamefont {K.~I.}\ \bibnamefont
  {Mazzitello}}, \bibinfo {author} {\bibfnamefont {J.}~\bibnamefont {Candia}},
  \ and\ \bibinfo {author} {\bibfnamefont {E.~V.}\ \bibnamefont {Albano}},\
  }\href@noop {} {\bibfield  {journal} {\bibinfo  {journal} {Physical Review
  E}\ }\textbf {\bibinfo {volume} {91}},\ \bibinfo {pages} {042118} (\bibinfo
  {year} {2015})}\BibitemShut {NoStop}%
\bibitem [{\citenamefont {Hagan}\ and\ \citenamefont
  {Chandler}(2006)}]{hagan2006dynamic}%
  \BibitemOpen
  \bibfield  {author} {\bibinfo {author} {\bibfnamefont {M.~F.}\ \bibnamefont
  {Hagan}}\ and\ \bibinfo {author} {\bibfnamefont {D.}~\bibnamefont
  {Chandler}},\ }\href@noop {} {\bibfield  {journal} {\bibinfo  {journal}
  {Biophysical Journal}\ }\textbf {\bibinfo {volume} {91}},\ \bibinfo {pages}
  {42} (\bibinfo {year} {2006})}\BibitemShut {NoStop}%
\bibitem [{\citenamefont {Wilber}\ \emph {et~al.}(2007)\citenamefont {Wilber},
  \citenamefont {Doye}, \citenamefont {Louis}, \citenamefont {Noya},
  \citenamefont {Miller},\ and\ \citenamefont {Wong}}]{wilber2007reversible}%
  \BibitemOpen
  \bibfield  {author} {\bibinfo {author} {\bibfnamefont {A.~W.}\ \bibnamefont
  {Wilber}}, \bibinfo {author} {\bibfnamefont {J.~P.}\ \bibnamefont {Doye}},
  \bibinfo {author} {\bibfnamefont {A.~A.}\ \bibnamefont {Louis}}, \bibinfo
  {author} {\bibfnamefont {E.~G.}\ \bibnamefont {Noya}}, \bibinfo {author}
  {\bibfnamefont {M.~A.}\ \bibnamefont {Miller}}, \ and\ \bibinfo {author}
  {\bibfnamefont {P.}~\bibnamefont {Wong}},\ }\href@noop {} {\bibfield
  {journal} {\bibinfo  {journal} {The Journal of Chemical Physics}\ }\textbf
  {\bibinfo {volume} {127}},\ \bibinfo {pages} {085106} (\bibinfo {year}
  {2007})}\BibitemShut {NoStop}%
\bibitem [{\citenamefont {Rapaport}(2008)}]{rapaport2008role}%
  \BibitemOpen
  \bibfield  {author} {\bibinfo {author} {\bibfnamefont {D.}~\bibnamefont
  {Rapaport}},\ }\href@noop {} {\bibfield  {journal} {\bibinfo  {journal}
  {Physical Review Letters}\ }\textbf {\bibinfo {volume} {101}},\ \bibinfo
  {pages} {186101} (\bibinfo {year} {2008})}\BibitemShut {NoStop}%
\bibitem [{\citenamefont {Drossel}\ and\ \citenamefont
  {Kardar}(1997)}]{drossel1997model}%
  \BibitemOpen
  \bibfield  {author} {\bibinfo {author} {\bibfnamefont {B.}~\bibnamefont
  {Drossel}}\ and\ \bibinfo {author} {\bibfnamefont {M.}~\bibnamefont
  {Kardar}},\ }\href@noop {} {\bibfield  {journal} {\bibinfo  {journal}
  {Physical Review E}\ }\textbf {\bibinfo {volume} {55}},\ \bibinfo {pages}
  {5026} (\bibinfo {year} {1997})}\BibitemShut {NoStop}%
\bibitem [{\citenamefont {Candia}\ and\ \citenamefont
  {Albano}(2008)}]{candia2008magnetic}%
  \BibitemOpen
  \bibfield  {author} {\bibinfo {author} {\bibfnamefont {J.}~\bibnamefont
  {Candia}}\ and\ \bibinfo {author} {\bibfnamefont {E.~V.}\ \bibnamefont
  {Albano}},\ }\href@noop {} {\bibfield  {journal} {\bibinfo  {journal}
  {International Journal of Modern Physics C}\ }\textbf {\bibinfo {volume}
  {19}},\ \bibinfo {pages} {1617} (\bibinfo {year} {2008})}\BibitemShut
  {NoStop}%
\bibitem [{\citenamefont {Candia}\ and\ \citenamefont
  {Albano}(2001)}]{candia2001comparative}%
  \BibitemOpen
  \bibfield  {author} {\bibinfo {author} {\bibfnamefont {J.}~\bibnamefont
  {Candia}}\ and\ \bibinfo {author} {\bibfnamefont {E.~V.}\ \bibnamefont
  {Albano}},\ }\href@noop {} {\bibfield  {journal} {\bibinfo  {journal}
  {Physical Review E}\ }\textbf {\bibinfo {volume} {63}},\ \bibinfo {pages}
  {066127} (\bibinfo {year} {2001})}\BibitemShut {NoStop}%
\bibitem [{\citenamefont {Gillespie}(2005)}]{gillespie2005general}%
  \BibitemOpen
  \bibfield  {author} {\bibinfo {author} {\bibfnamefont {D.}~\bibnamefont
  {Gillespie}},\ }\href@noop {} {\bibfield  {journal} {\bibinfo  {journal}
  {Journal of Computational Physics}\ }\textbf {\bibinfo {volume} {22}},\
  \bibinfo {pages} {403} (\bibinfo {year} {2005})}\BibitemShut {NoStop}%
\bibitem [{\citenamefont {Morris}\ and\ \citenamefont
  {Rogers}(2014)}]{morris2014growth}%
  \BibitemOpen
  \bibfield  {author} {\bibinfo {author} {\bibfnamefont {R.~G.}\ \bibnamefont
  {Morris}}\ and\ \bibinfo {author} {\bibfnamefont {T.}~\bibnamefont
  {Rogers}},\ }\href {http://stacks.iop.org/1751-8121/47/i=34/a=342003}
  {\bibfield  {journal} {\bibinfo  {journal} {Journal of Physics A:
  Mathematical and Theoretical}\ }\textbf {\bibinfo {volume} {47}},\ \bibinfo
  {pages} {342003} (\bibinfo {year} {2014})}\BibitemShut {NoStop}%
\bibitem [{\citenamefont {Chandler}(1987)}]{chandler}%
  \BibitemOpen
  \bibfield  {author} {\bibinfo {author} {\bibfnamefont {D.}~\bibnamefont
  {Chandler}},\ }\href@noop {} {\emph {\bibinfo {title} {{Introduction to
  modern statistical mechanics}}}}\ (\bibinfo  {publisher} {Oxford University
  Press New York},\ \bibinfo {year} {1987})\BibitemShut {NoStop}%
\bibitem [{\citenamefont {Binney}\ \emph {et~al.}(1992)\citenamefont {Binney},
  \citenamefont {Dowrick}, \citenamefont {Fisher},\ and\ \citenamefont
  {Newman}}]{binney1992theory}%
  \BibitemOpen
  \bibfield  {author} {\bibinfo {author} {\bibfnamefont {J.~J.}\ \bibnamefont
  {Binney}}, \bibinfo {author} {\bibfnamefont {N.}~\bibnamefont {Dowrick}},
  \bibinfo {author} {\bibfnamefont {A.}~\bibnamefont {Fisher}}, \ and\ \bibinfo
  {author} {\bibfnamefont {M.}~\bibnamefont {Newman}},\ }\href@noop {} {\emph
  {\bibinfo {title} {The theory of critical phenomena: an introduction to the
  renormalization group}}}\ (\bibinfo  {publisher} {Oxford University Press,
  Inc.},\ \bibinfo {year} {1992})\BibitemShut {NoStop}%
\bibitem [{Note1()}]{Note1}%
  \BibitemOpen
  \bibinfo {note} {The same conclusion follows from the variant of Eq. (14) of
  Ref.~\cite {morris2014growth} that would be obtained by setting, in Eq. (5)
  of that reference, the spin-flip terms $f$ to zero and requiring the
  vanishing of the remaining term.}\BibitemShut {Stop}%
\bibitem [{\citenamefont {Grinstein}\ \emph {et~al.}(1985)\citenamefont
  {Grinstein}, \citenamefont {Jayaprakash},\ and\ \citenamefont
  {He}}]{grinstein1985statistical}%
  \BibitemOpen
  \bibfield  {author} {\bibinfo {author} {\bibfnamefont {G.}~\bibnamefont
  {Grinstein}}, \bibinfo {author} {\bibfnamefont {C.}~\bibnamefont
  {Jayaprakash}}, \ and\ \bibinfo {author} {\bibfnamefont {Y.}~\bibnamefont
  {He}},\ }\href@noop {} {\bibfield  {journal} {\bibinfo  {journal} {Physical
  Review Letters}\ }\textbf {\bibinfo {volume} {55}},\ \bibinfo {pages} {2527}
  (\bibinfo {year} {1985})}\BibitemShut {NoStop}%
\bibitem [{\citenamefont {Bar-Yam}\ \emph {et~al.}(1990)\citenamefont
  {Bar-Yam}, \citenamefont {Kandel},\ and\ \citenamefont
  {Domany}}]{bar1990structure}%
  \BibitemOpen
  \bibfield  {author} {\bibinfo {author} {\bibfnamefont {Y.}~\bibnamefont
  {Bar-Yam}}, \bibinfo {author} {\bibfnamefont {D.}~\bibnamefont {Kandel}}, \
  and\ \bibinfo {author} {\bibfnamefont {E.}~\bibnamefont {Domany}},\
  }\href@noop {} {\bibfield  {journal} {\bibinfo  {journal} {Physical Review
  B}\ }\textbf {\bibinfo {volume} {41}},\ \bibinfo {pages} {12869} (\bibinfo
  {year} {1990})}\BibitemShut {NoStop}%
\bibitem [{\citenamefont {Whitelam}\ \emph {et~al.}(2014)\citenamefont
  {Whitelam}, \citenamefont {Hedges},\ and\ \citenamefont
  {Schmit}}]{whitelam2014self}%
  \BibitemOpen
  \bibfield  {author} {\bibinfo {author} {\bibfnamefont {S.}~\bibnamefont
  {Whitelam}}, \bibinfo {author} {\bibfnamefont {L.~O.}\ \bibnamefont
  {Hedges}}, \ and\ \bibinfo {author} {\bibfnamefont {J.~D.}\ \bibnamefont
  {Schmit}},\ }\href@noop {} {\bibfield  {journal} {\bibinfo  {journal}
  {Physical Review Letters}\ }\textbf {\bibinfo {volume} {112}},\ \bibinfo
  {pages} {155504} (\bibinfo {year} {2014})}\BibitemShut {NoStop}%
\bibitem [{\citenamefont {Whitelam}\ \emph {et~al.}(2016)\citenamefont
  {Whitelam}, \citenamefont {Dahal},\ and\ \citenamefont
  {Schmit}}]{whitelam2015minimal}%
  \BibitemOpen
  \bibfield  {author} {\bibinfo {author} {\bibfnamefont {S.}~\bibnamefont
  {Whitelam}}, \bibinfo {author} {\bibfnamefont {Y.~R.}\ \bibnamefont {Dahal}},
  \ and\ \bibinfo {author} {\bibfnamefont {J.~D.}\ \bibnamefont {Schmit}},\
  }\href@noop {} {\bibfield  {journal} {\bibinfo  {journal} {The Journal of
  Chemical Physics}\ }\textbf {\bibinfo {volume} {144}},\ \bibinfo {eid}
  {064903} (\bibinfo {year} {2016})}\BibitemShut {NoStop}%
\bibitem [{\citenamefont {Hedges}\ \emph {et~al.}(2009)\citenamefont {Hedges},
  \citenamefont {Jack}, \citenamefont {Garrahan},\ and\ \citenamefont
  {Chandler}}]{hedges2009dynamic}%
  \BibitemOpen
  \bibfield  {author} {\bibinfo {author} {\bibfnamefont {L.~O.}\ \bibnamefont
  {Hedges}}, \bibinfo {author} {\bibfnamefont {R.~L.}\ \bibnamefont {Jack}},
  \bibinfo {author} {\bibfnamefont {J.~P.}\ \bibnamefont {Garrahan}}, \ and\
  \bibinfo {author} {\bibfnamefont {D.}~\bibnamefont {Chandler}},\ }\href@noop
  {} {\bibfield  {journal} {\bibinfo  {journal} {Science}\ }\textbf {\bibinfo
  {volume} {323}},\ \bibinfo {pages} {1309} (\bibinfo {year}
  {2009})}\BibitemShut {NoStop}%
\bibitem [{\citenamefont {Sear}(2007)}]{sear2007nucleation}%
  \BibitemOpen
  \bibfield  {author} {\bibinfo {author} {\bibfnamefont {R.~P.}\ \bibnamefont
  {Sear}},\ }\href@noop {} {\bibfield  {journal} {\bibinfo  {journal} {Journal
  of Physics: Condensed Matter}\ }\textbf {\bibinfo {volume} {19}},\ \bibinfo
  {pages} {033101} (\bibinfo {year} {2007})}\BibitemShut {NoStop}%
\bibitem [{\citenamefont {Redner}\ \emph {et~al.}(2016)\citenamefont {Redner},
  \citenamefont {Wagner}, \citenamefont {Baskaran},\ and\ \citenamefont
  {Hagan}}]{redner2016classical}%
  \BibitemOpen
  \bibfield  {author} {\bibinfo {author} {\bibfnamefont {G.~S.}\ \bibnamefont
  {Redner}}, \bibinfo {author} {\bibfnamefont {C.~G.}\ \bibnamefont {Wagner}},
  \bibinfo {author} {\bibfnamefont {A.}~\bibnamefont {Baskaran}}, \ and\
  \bibinfo {author} {\bibfnamefont {M.~F.}\ \bibnamefont {Hagan}},\ }\href@noop
  {} {\bibfield  {journal} {\bibinfo  {journal} {arXiv preprint
  arXiv:1603.01362}\ } (\bibinfo {year} {2016})}\BibitemShut {NoStop}%
\end{thebibliography}

%

\onecolumngrid
\clearpage

\renewcommand{\theequation}{S\arabic{equation}}
\renewcommand{\thefigure}{S\arabic{figure}}
\renewcommand{\thesection}{S\arabic{section}}

\setcounter{equation}{0}
\setcounter{section}{0}
\setcounter{figure}{0}

\setlength{\parskip}{0.25cm}%
\setlength{\parindent}{0pt}%

\section{Stochastic models of growth}
\label{supp1}

The stochastic processes described in the main text can be described using a master equation. Consider the probability $P(r,b;t)$ that a system at time $t$ contains $r \geq 0 $ red and $b \geq 0$ blue particles. For brevity we will write this probability as $P(r,b)$, with the time-dependence of the function being implicit. We add red particles to the system with rate $\lr(r,b)$ and blue particles with rate $\lb(r,b)$, and remove red and blue particles with respective rates $\gr(r,b)$ and $\gb(r,b)$. The master equation for this set of processes is
\bea
\label{master}
\partial_t P(r,b) &=& \lb(r,b-1) P(r,b-1)-\lb(r,b) P(r,b) \nonumber \\
&+& \lr(r-1,b) P(r-1,b)-\lr(r,b) P(r,b) \nonumber \\
&+& \gb(r,b+1) P(r,b+1)-\gb(r,b) P(r,b) \nonumber \\
&+& \gr(r+1,b) P(r+1,b)-\gr(r,b) P(r,b).
\eea
We set $\gb(r,0) = 0 = \gr(0,b)$ so that we cannot have a negative number of particles of either color. By multiplying both sides of \eq{master} by the arbitrary function $U(r,b)$ and summing over $b$ and $r$ we obtain the evolution equation for the quantity $U$ averaged over dynamic trajectories, $\av{U(r,b)} \equiv \sum_{r,b=0}^\infty U(r,b) P(r,b)$:
\bea
\label{time_evolution_one}
\partial_t \langle U(r,b) \rangle &=& \langle \left[ U(r,b+1)-U(r,b) \right]\lb(r,b) \rangle \nonumber \\
&+&  \langle \left[ U(r+1,b)-U(r,b) \right] \lr(r,b) \rangle \nonumber \\
&+& \langle \left[ U(r,b-1)-U(r,b) \right]\gb(r,b) \rangle \nonumber \\
&+&  \langle \left[ U(r-1,b)-U(r,b) \right]\gr(r,b) \rangle. 
\eea
Setting $U(r,b)=b$ gives the rate of change of the mean number of blue particles,
\beq
\label{blue_ev}
\partial_t \av{b} = \av{\lb(b,r)}-\av{\gb(b,r)}.
\eeq
The corresponding equation for red particles is
\beq
\label{red_ev}
\partial_t \av{r} = \av{\lr(b,r)}-\av{\gr(b,r)}.
\eeq
We can obtain closed-form equations for rates of change of particle number by making a mean-field approximation, replacing averages over functions $f$ of $r$ and $b$ with functions $f$ of the averages of $r$ and $b$, i.e. writing $\av{f(r,b)} = f(\av{r},\av{b})$. To simplify notation we then replace $\av{r} \to r$ and $\av{b} \to b$, so that \eq{blue_ev} and \eq{red_ev} read
\bea
\label{one}
\dot{b} &=& \lb(b,r)-\gb(b,r);\\
\label{two}
\dot{r} &=& \lr(b,r)-\gr(b,r).
\eea
The size of the system is $N = r+b$, and so its growth rate is
\bea
\label{growth_rate}
\dot{N} &=& \dot{r} + \dot{b} \nonumber \\
&=& \lb+\lr-\gb-\gr.
\eea
In equilibrium we must have the vanishing of \eq{one} and \eq{two}, giving
\beq
\label{eos1}
\lb=\gb
\eeq
and
\beq
\label{eos2}
\lr=\gr.
\eeq
The rate of change of magnetization $m \equiv (b-r)/(b+r)$ is
\bea
\label{mdot}
\dot{m} &=& \frac{1}{N} \left[ \dot{b} - \dot{r} - m(\dot{b}+\dot{r})\right] \nonumber \\
&=& \frac{1}{N} \left[\lb-\lr-\gb+\gr-m(\lb+\lr-\gr-\gb) \right].
\eea
The condition $\dot{m}=0$ implies
\beq
\label{eos_noneq}
m_\star=\frac{\lb-\lr-\gb+\gr}{\lb+\lr-\gr-\gb},
\eeq
in which all rates are understood to be evaluated at $m=m_\star$.

\section{Irreversible model of growth}
\label{supp2}


The irreversible model described in the main text allows no particle removal, $\gb(r,b) = 0 = \gr(r,b)$. Blue particles are added with an Arrhenius-like rate that assumes Ising-like interaction interaction energies between red and blue particles with coupling $J$ and magnetic field $h$ (we take $\kt = 1$),
\bea
\lb(b,r) &=&\frac{1}{2} \exp\left\{ J  \frac{1+m}{2}-  J  \frac{1-m}{2}+ h  \right\} \nonumber \\
&=&\frac{1}{2} \e^{m J +h}.
\eea
 Here the spatial mean-field approximation is apparent, i.e. particles `feel' only the average magnetization of the whole system. We have absorbed the particle coordination number, assumed to be constant, into $J$. Similarly, red particles are added to the system with rate
\bea
\lr(b,r) &=&\frac{1}{2} \exp\left\{  J  \frac{1-m}{2}-  J  \frac{1+m}{2}- h  \right\} \nonumber \\
&=&\frac{1}{2} \e^{-m J- h},
\eea
The averaged growth rate \eq{growth_rate} is
\beq
\label{eden1}
\dot{N} = \cosh(m J +  h).
\eeq
The averaged time evolution of the system's magnetization, \eq{mdot}, is
\bea
\label{eden2}
\dot{m} &=& N^{-1} \left[ \sinh(m J + h) -m \cosh(m J +  h)\right].
\eea
This vanishes for
\beq
\label{eden3}
m_\star = \tanh(m_\star J +  h).
\eeq
Equations \eq{eden2} and \eq{eden3} are equations \eq{mdot_eden} and \eq{eos_eden} of the main text.

The temporal evolution to the attractor \eq{eden3} differs in different regimes of parameter space. Consider the case of vanishing magnetic field. For a small departure $\delta m$ from the attractor, $m(t)  = m_\star + \delta m(t)$, we have from \eq{eden1} $N \approx \cosh (m_\star J)t$. Inserting this result into \eq{eden2} and using \eq{eden3} gives
\beq
\partial_t \delta m \approx -\frac{1}{t} \left\{ \delta m \cosh(J \delta m)+[(m_\star)^2+m_\star \delta m-1] \sinh(J \delta m) \right\}.
\eeq
Expanding this equation in powers of $\delta m$ gives
\beq
\label{expansion}
\partial_t \delta m \approx \frac{1}{t} \left\{ \left[J(1-(m_\star)^2)-1\right] \delta m - J m_\star (\delta m)^2 +\frac{J^2}{6} \left[J(1-(m_\star)^2)-3\right] (\delta m)^3 \right\}.
\eeq
In the unmagnetized region $(J<1)$ we have $m_\star = 0$, and so $ \partial_t \delta m \approx (J-1) \delta m/t$. Thus temporal relaxation to the attractor is algebraic, with a continuously varying exponent: $\delta m \sim t^{J-1}$. In the magnetized region $(J>1,m_\star \neq 0)$ relaxation to steady-state is also algebraic, $\delta m \sim t^{J(1-(m_\star)^2)-1}$. Close to the critical point, where $J \approx 1$, we have from \eq{eden3} that $(m_\star)^2 \approx 3(J-1)/J^3$, and so $\delta m \sim t^{2(1-J)}$ to leading order in $J-1$. Thus the (moduli of) exponents either side of the critical point are distinct. At the critical point we have $J=1$ and $m_\star=0$, in which case the first two terms on the right-hand side of \eq{expansion} vanish. We then have $\partial_t \delta m \propto - (\delta m)^3/t$, and so $\delta m \sim (\ln t)^{-1/2}$.

\section{Reversible model of growth}
\label{supp3}

For the reversible model we have constant rates of particle addition, $\lb(r,b) = p c $ and $\lr(r,b) = (1-p) c$, where $c$ is a notional `solution' concentration and $p$ is the fraction of particles in solution that are blue. To make contact with Ising model nomenclature we introduce the magnetic field $h$ via $p \equiv \e^h/(2 \cosh h)$. Particle unbinding rates are Arrhenius-like, appropriate for particles escaping from a structure via themal fluctuations (we take $\kt = 1$):
\bea
\gb(b,r) &=& \frac{1+m}{2} \exp\left\{ -\es \frac{1+m}{2} - \ed \frac{1-m}{2} \right\} \nonumber \\
&\equiv&\frac{1+m}{2} \e^{-\Sigma -m \Delta},
\eea
where the magnetization of the system is again $m\equiv (b-r)/(b+r)$. We assume that blue particles possess energy of interaction $-\es$ with blue particles and $-\ed$ with red particles (we have absorbed factors of coordination number, assumed to be constant, into these energetic parameters). We have defined the parameters $\Sigma \equiv \left(\es+\ed \right)/2$ and $\Delta \equiv  \left(\es-\ed \right)/2$. The prefactor of the exponential ensures that blue particles leave the system with a rate proportional to their relative abundance in the system. For red particles we choose the unbinding rate
\bea
\gr(b,r) &=& \frac{1-m}{2} \exp\left\{ -\es \frac{1-m}{2} - \ed \frac{1+m}{2} \right\} \nonumber \\
&\equiv&\frac{1-m}{2} \e^{-\Sigma +m \Delta}.
\eea
Note that because $m$ is not defined for $r=b=0$ we additionally require $\gb(0,0) = 0 = \gr(0,0)$. 

Hence \eq{one} and \eq{two} become
\bea
\dot{b}&=&pc-\frac{1+m}{2} \e^{-\Sigma - m \Delta},\\
\dot{r}&=&(1-p)c-\frac{1-m}{2} \e^{-\Sigma + m \Delta},
\eea
which, with $p=1/2$, are Equations (1) of Ref.\c{whitelam2014self}. It is convenient to rescale time and concentration
\beq
t \to \e^{\Sigma} t
\eeq
and
\beq
c \to \e^{-\Sigma} c
\eeq
to get the simpler equations
\bea
\dot{b}&=&pc-\frac{1+m}{2} \e^{- m \Delta}, \label{rate1}\\
\dot{r}&=&(1-p)c-\frac{1-m}{2} \e^{ m \Delta}. \label{rate2}
\eea
The growth rate \eq{growth_rate} is
\beq
\label{ndot_supp}
\dot{N} = c-  \cosh(m \Delta) +m  \sinh (m \Delta).
\eeq
In this model there exists an equilibrium when rates of particle addition and removal balance. The the associated equation of state follows from \eq{eos1} and \eq{eos2}, and is
\beq
\label{eq1_supp}
m_0 = \tanh(m_0 \Delta+h)
\eeq
with the associated concentration
\beq
\label{eq2}
c_0^2 = \left( 1-m_0^2\right) \cosh^2 h.
\eeq
Note that the equilibrium concentration for $h=0$ is the same for red $(m_0<0)$ and blue $(m_0>0)$ solutions, i.e. $c_0$ is unchanged upon setting $m_0 \to -m_0$. 

The rate of magnetization evolution, \eqq{mdot}, is 
\beq
\label{mdot2_supp}
\dot{m} = \frac{1}{N} \left[ \left(1-m^2\right) \sinh(m \Delta) - c \left(m-\tanh h\right)\right],
\eeq
which vanishes when
\beq
\label{ss_m_supp}
c\left(m_\star -\tanh h\right)=\left( 1-(m_\star)^2\right) \sinh (m_\star \Delta).
\eeq

In the main text we assume an Ising-like hierarchy for the interaction energies of this model, in which case $\Delta = J$ and $\Sigma=0$. With these choices Equations~\eq{eq1_supp}, \eq{mdot2_supp}, and \eq{ss_m_supp} are equations \eq{rev_3}, \eq{rev_2}, and \eq{eos_rev} of the main text

Analysis of \eq{ss_m_supp}, for $h=0$, gives rise to \f{fig1}(a) of the main text. The function on the right-hand side of \eq{ss_m_supp} vanishes at $m = 0$ and at $m= \pm 1$, and has one turning point for positive $m$ and one for negative $m$. When $\Delta < \sqrt{6}$ this function takes its largest positive gradient, $\Delta$, at the origin. Therefore it intersects the function $c m$ on the left-hand side of \eq{ss_m_supp} three times if $c< \Delta$ (with two non-negative solutions, $m_{\pm}$, stable to perturbations of $m$, and one, at $m=0$, unstable) and only once (at $m=0$) if $c> \Delta$. When $c=\Delta$ all solutions lie at $m=0$. The solutions $m_{\pm}$ move smoothly away from $m$ as $c$ is decreased below $\Delta$, and do so as
\beq
\label{scaling}
m_\star \sim \left( \frac{6}{\Delta}\cdot\frac{\Delta-c}{6-\Delta^2}\right)^{1/2}.
\eeq
Thus at the point $c=\Delta$ (for $h=0$ and $\Delta < \sqrt{6}$) we have a continuous nonequilibrium phase transition separating zero- and nonzero magnetization solutions to \eqq{ss_m_supp}.

For $h=0$ and $\Delta \geq \sqrt{6}$ we can have zero, three or five solutions to \eq{ss_m_supp}, depending on the value of $c$, and respectively zero, two and three of those solutions are stable. 

As for the irreversible model, temporal relaxation to the attractor $m_\star$ varies by parameter regime. Expanding \eq{ndot_supp} and \eq{mdot2_supp} for $m(t)=m_\star+\delta m(t)$ gives, for $m_\star=0$,
\beq
\partial_t \delta m \approx \frac{1}{t} \left\{ \frac{\Delta-c}{c-1} \delta m - \frac{\Delta (6-\Delta^2)}{6(\Delta -1)} (\delta m)^3 \right\}.
\eeq 
In the unmagnetized region $c>\Delta$ we then have $\delta m \sim t^{\frac{\Delta-c}{c-1}}$. At criticality $(\Delta = c)$ we have $\delta m \sim (\ln t)^{-1/2}$. Expanding \eq{ndot_supp} and \eq{mdot2_supp} for $m(t)=m_\star+\delta m(t)$ and using \eq{scaling} reveals that in the magnetized region we have $\delta m \sim t^{\frac{2(c-\Delta)}{c-1}}$.

\end{document}